\documentstyle[twoside,psfig,epsf]{article}

\catcode`\@=11 \long\def\@makefntext#1{ \protect\noindent \hbox to
3.2pt {\hskip-.9pt
$^{{\eightrm\@thefnmark}}$\hfil}#1\hfill}       

\def\@makefnmark{\hbox to 0pt{$^{\@thefnmark}$\hss}}    

\def\ps@myheadings{\let\@mkboth\@gobbletwo
\def\@oddhead{\hfill\hbox{}\rightmark}
\def\@oddfoot{}\def\@evenhead{\leftmark\hbox{}\hfill}\def\@evenfoot{}
\def\sectionmark##1{}\def\subsectionmark##1{}}

\oddsidemargin=\evensidemargin 
\newcounter{sectionc}\newcounter{subsectionc}\newcounter{subsubsectionc}
\renewcommand{\section}[1] {\vspace{12pt}\addtocounter{sectionc}{1}
\setcounter{subsectionc}{0}\setcounter{subsubsectionc}{0}\noindent
    {\tenbf\thesectionc. #1}\par\vspace{5pt}}
\renewcommand{\subsection}[1] {\vspace{12pt}\addtocounter{subsectionc}{1}
    \setcounter{subsubsectionc}{0}\noindent
  {\bf\thesectionc.\thesubsectionc. {\kern1pt \bfit #1}}\par\vspace{5pt}}
\renewcommand{\subsubsection}[1] {\vspace{12pt}\addtocounter{subsubsectionc}{1}
    \noindent{\tenrm\thesectionc.\thesubsectionc.\thesubsubsectionc.
    {\kern1pt \tenit #1}}\par\vspace{5pt}}
\newcommand{\nonumsection}[1] {\vspace{12pt}\noindent{\tenbf #1}
    \par\vspace{5pt}}

\newcounter{appendixc}
\newcounter{subappendixc}[appendixc]
\newcounter{subsubappendixc}[subappendixc]
\renewcommand{\thesubappendixc}{\Alph{appendixc}.\arabic{subappendixc}}
\renewcommand{\thesubsubappendixc}
    {\Alph{appendixc}.\arabic{subappendixc}.\arabic{subsubappendixc}}

\renewcommand{\appendix}[1] {\vspace{12pt}
        \refstepcounter{appendixc}
        \setcounter{figure}{0}
        \setcounter{table}{0}
        \setcounter{lemma}{0}
        \setcounter{theorem}{0}
        \setcounter{corollary}{0}
        \setcounter{definition}{0}
        \setcounter{equation}{0}
        \renewcommand{\thefigure}{\Alph{appendixc}.\arabic{figure}}
        \renewcommand{\thetable}{\Alph{appendixc}.\arabic{table}}
        \renewcommand{\theappendixc}{\Alph{appendixc}}
        \renewcommand{\thelemma}{\Alph{appendixc}.\arabic{lemma}}
        \renewcommand{\thetheorem}{\Alph{appendixc}.\arabic{theorem}}
        \renewcommand{\thedefinition}{\Alph{appendixc}.\arabic{definition}}
        \renewcommand{\thecorollary}{\Alph{appendixc}.\arabic{corollary}}
        \renewcommand{\theequation}{\Alph{appendixc}.\arabic{equation}}
        \noindent{\tenbf Appendix#1}\par\vspace{5pt}}
\newcommand{\subappendix}[1] {\vspace{12pt}
        \refstepcounter{subappendixc}
        \noindent{\bf Appendix \thesubappendixc. {\kern1pt \bfit #1}}
    \par\vspace{5pt}}
\newcommand{\subsubappendix}[1] {\vspace{12pt}
        \refstepcounter{subsubappendixc}
        \noindent{\rm Appendix \thesubsubappendixc. {\kern1pt \tenit #1}}
    \par\vspace{5pt}}

\newcommand{\textlineskip}{\baselineskip=13pt}
\newcommand{\smalllineskip}{\baselineskip=10pt}

\newcommand{\copyrightheading}[1]
    {\vspace*{-2.5cm}\smalllineskip{\flushleft
    {\footnotesize Preprint submitted to International Journal of Theoretical and
     Applied Finance #1}\\
    {\footnotesize \copyright\kern2pt World Scientific Publishing
     Company}\\
     }}

\def\abstracts#1#2#3{{
    \centering{\begin{minipage}{4.5in}\footnotesize\baselineskip=10pt
    \parindent=0pt #1\par
    \parindent=15pt #2\par
    \parindent=15pt #3
    \end{minipage}}\par}}

\def\keywords#1{{
    \centering{\begin{minipage}{4.5in}\footnotesize\baselineskip=10pt
    {\footnotesize\it Keywords}\/: #1
     \end{minipage}}\par}}

\renewenvironment{thebibliography}[1]
    {\frenchspacing
     \ninerm\baselineskip=11pt
     \begin{list}{[\arabic{enumi}]}
    {\usecounter{enumi}\setlength{\parsep}{0pt}
     \setlength{\leftmargin 15pt}{\rightmargin 0pt} 
     \setlength{\itemsep}{0pt} \settowidth
    {\labelwidth}{[#1]}\sloppy}}{\end{list}}

\newcounter{itemlistc}
\newcounter{romanlistc}
\newcounter{alphlistc}
\newcounter{arabiclistc}

\newcommand{\fcaption}[1]{
        \refstepcounter{figure}
        \setbox\@tempboxa = \hbox{\footnotesize Fig.~\thefigure. #1}
        \ifdim \wd\@tempboxa > 5in
           {\begin{center}
        \parbox{5in}{\footnotesize\smalllineskip Fig.~\thefigure. #1}
            \end{center}}
        \else
             {\begin{center}
             {\footnotesize Fig.~\thefigure. #1}
              \end{center}}
        \fi}

\newcommand{\tcaption}[1]{
        \refstepcounter{table}
        \setbox\@tempboxa = \hbox{\footnotesize Table~\thetable. #1}
        \ifdim \wd\@tempboxa > 5in
           {\begin{center}
        \parbox{5in}{\footnotesize\smalllineskip Table~\thetable. #1}
            \end{center}}
        \else
             {\begin{center}
             {\footnotesize Table~\thetable. #1}
              \end{center}}
        \fi}

\def\@citex[#1]#2{\if@filesw\immediate\write\@auxout
    {\string\citation{#2}}\fi
\def\@citea{}\@cite{\@for\@citeb:=#2\do
    {\@citea\def\@citea{,}\@ifundefined
    {b@\@citeb}{{\bf ?}\@warning
    {Citation `\@citeb' on page \thepage \space undefined}}
    {\csname b@\@citeb\endcsname}}}{#1}}

\newif\if@cghi
\def\cite{\@cghitrue\@ifnextchar [{\@tempswatrue
    \@citex}{\@tempswafalse\@citex[]}}
\def\citelow{\@cghifalse\@ifnextchar [{\@tempswatrue
    \@citex}{\@tempswafalse\@citex[]}}
\def\@cite#1#2{{$\null^{#1}$\if@tempswa\typeout
    {IJCGA warning: optional citation argument
    ignored: `#2'} \fi}}

\def\pmb#1{\setbox0=\hbox{#1}
    \kern-.025em\copy0\kern-\wd0
    \kern.05em\copy0\kern-\wd0
    \kern-.025em\raise.0433em\box0}

\def\fnm#1{$^{\mbox{\scriptsize #1}}$}
\def\fnt#1#2{\footnotetext{\kern-.3em
    {$^{\mbox{\scriptsize #1}}$}{#2}}}

\font\tenrm=cmr10 \font\tenit=cmti10 \font\tenbf=cmbx10
\font\bfit=cmbxti10 at 10pt \font\ninerm=cmr9 \font\nineit=cmti9
\font\ninebf=cmbx9 \font\eightrm=cmr8

\def\qed{\hbox{${\vcenter{\vbox{         
   \hrule height 0.4pt\hbox{\vrule width 0.4pt height 6pt
   \kern5pt\vrule width 0.4pt}\hrule height 0.4pt}}}$}}

\def\theequation{\thesectionc.\arabic{equation}}  

\newenvironment{itemlist1}          
        {\setcounter{itemlistc}{0}      
     \begin{list}{$\bullet$}        
    {\usecounter{itemlistc}         
     \leftmargin10pt           
     \setlength{\parsep}{0pt}
     \setlength{\itemsep}{0pt}     
    }}{\end{list}}

    {\setcounter{romanlistc}{0}     
     \begin{list}{$($\roman{romanlistc}$)$} 
    {\usecounter{romanlistc}        
     \leftmargin18pt 
     \setlength{\parsep}{0pt}
     \setlength{\itemsep}{0pt}  
     \settowidth{\labelwidth}{#1}
    }}{\end{list}}

    {\setcounter{enumii}{0}         
     \begin{list}{$($\alph{enumii}$)$}  
    {\usecounter{enumii}            
     \leftmargin18pt        
     \setlength{\parsep}{0pt}
     \setlength{\itemsep}{0pt}  
     \settowidth{\labelwidth}{#1}
    }}{\end{list}}

\begin{document}
\setlength{\textheight}{7.7truein}    

\markboth{\protect{\footnotesize\it The Feedback Effect of Hedging
in Portfolio Optimization}}{\protect{\footnotesize\it P.
HENRY-LABORD\`ERE}}

\normalsize\textlineskip \thispagestyle{empty}
\setcounter{page}{1}

\copyrightheading{}     

\vspace*{1in}

\centerline{\bf The Feedback Effect of Hedging in Portfolio
Optimization} \baselineskip=13pt \vspace*{0.37truein}

\centerline{\footnotesize P. HENRY-LABORD\`ERE} \baselineskip=12pt
\centerline{\footnotesize\it Theoretical Physics Group, The
Blackett Laboratory, Imperial College London} \baselineskip=10pt
\centerline{\footnotesize\it Prince Consort Road, London SW7 2AZ,
U.K.} \baselineskip=10pt \centerline{\footnotesize\it
phenry@lpt.ens.fr} \vspace*{0.225truein}

\vspace*{0.21truein}
\abstracts{In this short note, we will show
how to optimize the portfolio of a large trader whose hedging
strategy affects the price of his assets.}{}{}

\vspace*{10pt} \keywords{Illiquid Market, feedback effect,
stochastic optimization}


\vspace*{4pt}
\normalsize\baselineskip=13pt   
\section{Introduction}
\noindent Since the famous papers of Black-Scholes on option
pricing \cite{bla} and of Markowitz on portfolio optimization
\cite{mar} (for a review \cite{hull}), some progress has been made
in order to extend these results to a more realistic
arbitrage-free market model. As a reminder, the Black-Scholes
theory and the Markowitz's theory consist in following a (hedging)
strategy to decrease the risk of loss given a fixed amount of
return. Both theories are based on three important hypothesis
which are satisfied only to a certain extent:

\begin{itemlist1}

\item The traders can revise their decisions continuously in time.
This first hypothesis is not realistic for obvious reasons.
A major improvement was recently introduced by  Bouchaud-Sornette
\cite{bs} in their time-discrete model. They introduce an
elementary time $\tau$ after which a trader is able to revise his
decisions again. The optimal strategy is fixed by the minimization
of the risk defined by the variance of the portfolio. The
resulting risk is no longer zero and in the continuous-time limit
where $\tau$ goes to zero, one recovers the classical result of
Black-Scholes: the risk vanishes (one can show by using the Ito's
formalism that this is true for any definition of the risk in the
context of continuous-time log-normal price processes).

\item The price fluctuations obey a log-normal law. This second
hypothesis doesn't truthfully reflect the market. A data analysis
would rather suggest that the (uncorrelated) price increments
$\delta x_n$ at time $t=n \tau$ ( with $\tau$ about 15-30 minutes)
are well fitted with a truncated Levy law or a Student
distribution \cite{bou}. Both distributions exhibit large jumps
for each increment and are distributed with fat tails, thus far
from being a log-normal law. Moreover, the $\delta x_n$ are no
longer uncorrelated below the 15-30 minutes interval (incomplete
markets). Thus the Bouchaud-Sornette model goes beyond the
classical financial theory and allows to incorporate non
log-normal price processes such as a truncated Levy law. If one
tries to reproduce the price of an option given by the
Bouchaud-Sornette theory by using the Black-Scholes model, one
must introduce an implied volatility which will depend on the
kurtosis of the non log-normal law. This theoretical implied
volatility then reproduces  the experimental "smile law"
\cite{bou}.

\item Markets are assumed to be completely elastic. This third
hypothesis means that small traders don't modify the prices of the
market by selling or buying large amounts of assets. The
limitation of this last assumption lies with the fact it is only
justified when the market is liquid.
\end{itemlist1}

\noindent  While so far research has mainly focused on relaxing
the first and/or the second hypothesis described above, in this
paper we shall relax only the third one. In relation to this, we
note that some recent articles have appeared in the case of the
hedging of derivatives in markets which are not perfectly liquid
\cite{wil,bre,fre,pla,gen}. In this context, the classical
Black-Scholes equation is  replaced by a non-linear partial
differential equation (PDE) which we give (in the case of our
analysis) in the appendix.
\\
\\
\noindent The aim of this paper is to analyze the feedback effect
of hedging in portfolio optimization. In the first section, we
model the market as composed of small traders and a large one
whose demand is given by a hedging strategy. We derive from this
toy-model the parameters of the stochastic process of the price
with the influence of the large trader. Taking into account the
influence on the volatility and the return, we compute in the
second section the hedging strategy of the large trader in order
to optimize dynamically a portfolio composed of a risky asset and
a bond. This leads to a well-defined stochastic optimization
problem.

\section{The feedback effect of hedging on price}
\noindent By definition, a market is liquid when the elasticity
parameter is small. The elasticity parameter $\epsilon$ is given
by the ratio of relative change in price $S_t$ to change in the
net demand $D$:
\begin{equation}  { dS_t \over S_t} = \epsilon dD_t \label{ela}\end{equation}
\noindent We observe experimentally that when the demand increases
(resp. decreases), the price rises (decreases). The parameter
$\epsilon$ is therefore positive and we will assume that it is
also constant. We will then reproduce the log-normal process when
the feedback hedging effect is negligible. Another interesting
(because more realistic) assumption would be to define $\epsilon$
as a stochastic variable.

\noindent In a crash situation, the small traders who tend to
apply the same hedging strategy can considered as a large trader
and the hedging feedback effects become very important. This  can
speed up the crash. In the following, we will use the simple
relation (\ref{ela}) to analyze the influence of dynamical trading
strategies on the prices in financial markets. The dynamical
trading $\phi(S_t,t)$ which represents the number of shares of a
given stock that a large trader holds will be determined in order
to optimize his portfolio.  An analytic solution will not be
possible and we will resolve the non-linear equations by expanding
the solution as a formal series in the parameter $\epsilon$. The
first order correction will be obtained.

\noindent Let's call $D(t,W,S_t)$ the demand of all the traders in
the market which depends on time $t$, a Brownian variable $W$ and
the price $S_t$. $S_t$ is assumed to satisfy the stochastic
equation
\begin{eqnarray}
{dS_t \over S_t}=\mu(t,S_t)dt+\sigma(t,S_t)dW \label{logn}
\end{eqnarray}
with $\mu(S_t,t)$ the return and $\sigma(S_t,t)$ the volatility.
The stochastic process $W$ models\- the information the traders
have on the demand. If we take the derivative of $D(t,S_t,W)$, ie
$dD$, we obtain (using the Ito's symbolic "rules" $dW^2 = dt$,
$dWdS=\sigma S dt$ and $dS^2=\sigma^2 S^2dt$)
\begin{eqnarray}
dD&=&(\partial_t D  + {1 \over 2} \sigma^2 S^2
\partial^2_{S} D   + {1 \over 2}\partial^2_{W} D  + \sigma S
\partial_{SW} D)  dt +\partial_{S} D dS_t +
\partial_{W} D dW  \label{der} \\
&=&{1 \over \epsilon} {dS_t \over S_t}={1 \over
\epsilon}(\mu(t,S_t)dt+\sigma(t,S_t)dW)
\end{eqnarray}
\noindent By identifying the coefficients for $dt$ and $dW$ in the
above stochastic equation, one obtains $\mu$ and $\sigma$ as a
function of the derivatives of $D$:
\begin{eqnarray}
\sigma(t,S_t)&=&{\epsilon \partial_W D \over (1-\epsilon
S\partial_S D)} \label{vold}
\\
\mu(t,S_t)&=&\epsilon[\partial_t D +{1 \over 2}\sigma^2 S^2
\partial_S^2 D +\sigma S \partial_{SW}D +{1 \over 2} \partial_W^2D] \over (1-\epsilon
S \partial_S D) \label{retd}
\end{eqnarray}
We will assume that the market is composed of a group of small
traders who don't modify the prices of the market by selling or
buying large amounts of assets on the one hand and a large trader
one the other hand. One can consider a large trader as an
aggregate of small traders following the same strategy given by
the hedging position $\phi$.

\noindent We will choose the demand $D_{small}$ of small traders
in order to reproduce the classic log-normal random walk motion
(\ref{logn}) with a constant return $\mu=\mu_0$ and a constant
volatility $\sigma=\sigma_0$. The simple solution is given by
\begin{equation} D_{{\mathrm small}}={1 \over \epsilon} (\mu_0 t+ \sigma_0W)
\end{equation}

\noindent Now, we include the effect of a large trader whose
demand $D_{{\mathrm large}}$ is generated by the trading strategy
$D_{{\mathrm large}}= \phi(t,S_t)$. It is implicitly assumed that
$\phi$ depends only on $S$ and $t$ and not on the history of the
Brownian motion. This hypothesis is proven in the appendix in the
case of option pricing by using a Black-Scholes analysis.
Moreover, we will see in the following section that this
hypothesis is not strictly valid for our portfolio trading
strategy. Indeed, $\phi$ will depend on $S$, $t$ and also the
value of our portfolio $\Pi$. This last value depends implicitly
on the history of the Brownian motion. To simplify the discussion,
we will take the mean value of the portfolio $\Pi$ in order to
obtain a trading strategy which will depend only on $S$ and $t$.

\noindent The hedging position $\phi$ is then added to the demand
of the small traders $D_{{\mathrm small}}$ and the total net
demand is  given by
\begin{equation}
D=D_{{\mathrm small}}+ \phi(t,S_t) \label{dem}
\end{equation}
By inserting (\ref{dem}) in (\ref{vold})-(\ref{retd}), one finds
the new volatility and return as a function of $\phi$:
\begin{eqnarray}
\sigma&=& { \sigma_0 \over (1-\epsilon S\partial_S \phi)} \label{vol}\\
\mu&=&{\mu_0+{\epsilon}(\partial_t \phi+ {1 \over 2}\sigma^2 S^2
\partial_S^2 \phi) \over (1-\epsilon S\partial_S \phi)}
\label{ret}
\end{eqnarray}
\noindent This relation describes the feedback effect of dynamical
hedging on volatility and return. The hypothesis that the hedging
function depends only on $S$ and $t$ allows to obtain a volatility
and a return independent of historic effects. For derivatives and
portfolio hedging strategies, the feedback effect on the
volatility will not be the same.

\noindent Let's take the case of derivatives hedging first. If the
price rises, then $\partial_S \phi(S,t)>0$  meaning that  the
trader buys additional shares. For completeness, it is shown in
the appendix that the Black-Scholes relation $\phi=\partial_S
{\cal C}$ ( with $\cal C$ the value of an option) is still valid
and the Gamma derivative $\Gamma=\partial_S \phi(S,t)=\partial_S^2
{\cal C}$ is positive. The volatility $\sigma$ is then greater
than $\sigma_0$ and the price decreases \fnm{a}\fnt{a}{The
zero-order solution of (\ref{logn}) (in the parameter $\epsilon$)
is $S_t=S_0 e^{(\mu_0-{1 \over 2} \sigma^2_0)t+\sigma_0 W} $ and
$S_t$ decreases as $\sigma_0$ increases.}. A destabilizing effect
is then obtained.

\noindent On the other hand, in dynamical portfolio optimization,
we have the expression $\partial_S \phi(S,t)<0$ (as we will see in
the next section) according to which  a trader should sell stocks
when his price increases and buy more stocks when his price
decreases. As a result  $\sigma < \sigma_0$ and the price
increases. In this situation, a trader could buy large amounts of
shares at a price $S$ and the price would then move to a higher
price $S'$. By selling his shares, it would make a free-risk
profit $S'-S$ per share. The key feature that allows this
manipulation is that the price reacts with a delay that allows the
trader to buy at low price and sell at a higher price before the
price goes down. It is shown in \cite{jar} that to prevent
manipulation strategies, the price must not react with delays.

\noindent In \cite{wil}, a slightly different approach is used to
model the influence of the hedging on the volatility and the
return. We will show that this approach can lead to the same
result as ours for the volatility. One considers the difference
$\chi$ between the demand $D(t,S_t,W)$ and the supply ${\cal
S}(t,S_t,W)$ which is the total number of shares available in the
market:
\begin{equation} \chi(t,S_t,W)=D-{\cal S} \end{equation} \noindent It is clear that if $\chi$
increases (resp. decreases), the price will increase (resp.
decrease) and the equilibrium price is reached when $\chi=0$. By
taking the derivative of $\chi$ (d$\chi=0$, see (\ref{der})), one
can then express $\mu$ and $\sigma$ appearing in (\ref{logn}) as a
function of $\chi$:
\begin{eqnarray}
\sigma&=&-{\partial_W \chi \over S \partial_S \chi} \\
\mu&=&-{(\partial_t \chi +{1 \over 2}\sigma^2 S^2 \partial_S^2
\chi +{1 \over 2}\partial_W^2 \chi +\sigma S \partial_{SW} D)
\over S\partial_S D}
\end{eqnarray}
The $\chi$ for the small traders is then chosen in order to find a
constant return and volatility. One should note that the solution
is not unique. One can take (this is not the choice of \cite{wil})
\begin{equation} \chi_{small}(S,W,t)={1 \over \epsilon}((\mu_0-{1 \over
2}\sigma_0^2)t+\sigma_0W-ln(S)) \end{equation} \noindent If we
include a large trader with a hedging position $\phi(t,S_t)$, the
equilibrium equation is easily modified as \noindent
\begin{equation} \chi_{small}(t,S,W) +\phi(t,S)=0
\end{equation} \noindent The modified volatility $\sigma$  then reproduces precisely the formula
(\ref{vol}) and the return differs slightly from (\ref{ret}) and
is given by:
\begin{equation}
\mu={\mu_0+{1 \over 2} \sigma^2 + \epsilon( \partial_t \phi+ {1
\over 2} \sigma^2 S^2 \partial_S^2 \phi) \over (1-\epsilon S
\partial_S \phi)} \label{volw}
\end{equation}

\noindent By using a different $\chi_{small}$, \cite{wil} found
expressions for the volatility and the return which differ from
(\ref{vol})-(\ref{ret})-(\ref{volw}).

\noindent The parameters $\mu$, $\sigma$ can be written as a
formal series in $\epsilon$ and we will see in the following
section that this is also the case for the hedging function
$\phi=\sum_{i=0} \phi_i(S,t) \epsilon^i$. We give the first order
expansion of $\sigma$ (\ref{vol}) and $\mu$ (\ref{ret}) :
\begin{eqnarray}
\sigma&=&\sigma_0(1+\epsilon S \partial_S \phi) +o(\epsilon^2) \label{vola}\\
\mu&=&\mu_0+{\epsilon}(\partial_t \phi+ {1 \over 2}\sigma_0^2 S^2
\partial_S^2 \phi+\mu_0 S \partial_S \phi)+o(\epsilon^2)
\label{reta}
\end{eqnarray}
The equation above will be used throughout the next section.

\section{Optimized portfolio of a large trader}
\noindent Let's assume that a large trader holds at time $t$ a
certain number of shares $\phi(S_t,t)$ of which the price at time
$t$ is $S_t$. The price $S_t$ satisfies the stochastic
differential equation (\ref{logn}) which depends implicitly on the
hedging position $\phi(S_t,t)$ (see the volatility (\ref{vol}) and
 the return (\ref{ret})). The trader holds also a number $b_t$ of
bonds the price of which satisfies the following deterministic
equation $dB_t=r(t) B_tdt$ with $r(t)$ the interest rate of the
bond. The change in his portfolio $\Pi=\phi(S,t) S +b_t B_t$
during an elementary time $dt$ is then

\begin{eqnarray} d{\Pi}&=&\phi(t,S_t)dS_t +r(t) b_t B_t dt\label{ito3}
\\
&=&r(t) \Pi dt + \phi(t,S_t) S_t((\mu-r(t)) dt+\sigma dW)
\label{ito2}
\end{eqnarray}
\noindent One should note that $\phi$ can be negative (which is
equivalent to a short position). At a maturity date $T$, the value
of the portfolio ${\Pi}_T$ is
\begin{equation}
{\Pi}_T=\Pi_0+\int_{0}^{T} d\Pi
\end{equation}

\noindent The basic strategy of the trader is to find the optimal
strategy $\phi^*(S,t)$ so that the risk $\cal R$ is minimized for
a given fixed value of the profit $E[\Pi_T]={\cal G}$ with
$E[\cdot]$ the mean operator. By definition, the risk is given by
the variance of the portfolio $\Pi_T$:
\begin{equation}
{\cal R} \doteq E[\Pi^2_T]-E[\Pi_T]^2=E[\Pi^2_T -{\cal G}^2]
\end{equation}
\noindent It is clear that the result of this paper depends on the
above mentioned definition for the risk, its main advantage being
that it gives simple computations. The mean-variance portfolio
selection problem is then formulated as the following optimization
problem parameterized by $\cal G$:
\begin{equation}
{\mathrm \; \; min \; \;} E[ \Pi_T^2-{\cal G}^2]
\end{equation}
\noindent subject to
\begin{equation}
E[\Pi_T]={\cal G} \;,\; , (S(.),\Pi(.)) {\; \; \mathrm satisfying
\;\; (\ref{logn})-(\ref{ito2})}
\end{equation}
This problem is equivalent to the following one in which we have
introduced a Lagrange multiplier $\zeta$:
\begin{equation}
{\mathrm \; \; min \; \;} E[ \Pi_T^2-{\cal G}^2 +\zeta(\Pi_T-{\cal
G})]
\end{equation}
\noindent subject to
\begin{equation}
(S(.),\Pi(.)) {\; \; \mathrm satisfying \;\;
(\ref{logn})-(\ref{ito2})}
\end{equation}
To simplify, we will assume in the following  that the interest
rates are negligible (ie $r(t)=0$) although this hypothesis can be
easily relaxed. \noindent To derive the optimality equations, the
above problem can be restated in a dynamic programming form so
that the Bellman principle of optimality can be applied
\cite{kirk}-\cite{merton}. To do this, let's define
\begin{equation}
{\cal J}(t,S,\Pi)={\mathrm min} \; \;  E_t[ \Pi_T^2-{\cal
G}^2+\zeta(\Pi_T-{\cal G})]
\end{equation}
where $E_t$ is the conditional expectation operator at time $t$
with $S_t=S$ and $\Pi_t=\Pi$. Then, $J$ satisfies the famous
Hamilton-Jacobi-Bellman equation:
\begin{equation}
{Min}_{\phi,\zeta}\{ \partial_t{\cal J}+\mu S \partial_S{\cal J}+
{1 \over 2} \sigma^2 S^2 \partial_S^2 {\cal J} + \mu S \phi
\partial_\Pi {\cal J}+{1 \over 2} \phi^2 \sigma^2 S^2
\partial_\Pi^2 {\cal J} + \phi \sigma^2 S^2 \partial_S
\partial_\Pi {\cal J}=0\} \label{H}
\end{equation}
\noindent subject to
\begin{eqnarray}
{\cal J}(T,S,\Pi)&=&\Pi_T^2-{\cal G}^2 +\zeta(\Pi_T-{\cal G}) \\
&=&(\Pi_T+{\zeta \over 2})^2 -{\cal G}^2-\zeta {\cal G}-{\zeta^2
\over 4} \label{bH}
\end{eqnarray}

\noindent The equation (\ref{H}) is quite complicated due to the
explicit dependence of the volatility and the return in the
control $\phi$. A simple method to solve the H-J-B equation
(\ref{H}) perturbatively in the liquidity parameter $\epsilon$ is
to use a mean-field approximation well known in statistical
physics: one computes the optimal control $\phi^*(S,t)$ up to a
given order $\epsilon^i$, say $\phi^*_i(S,t)$. This gives a mean
volatility $\sigma_i(S,t)=\sigma(\phi^*_i(S,t),S,t)$ and mean
return $\mu_i(S,t)=\mu(\phi^*_i(S,t),S,t)$ up to a given order
$\epsilon^{i+1}$. The optimal control, at the order
$\epsilon^{i+1}$, is then easily found by taking the functional
derivative of the expression (\ref{H}) according to $\phi$:
\begin{equation}
\phi^*_{i+1}=-{(\mu_i \partial_\Pi {\cal J} +\sigma_i^2 S
\partial_{S \Pi} {\cal J}) \over \sigma_i^2 S \partial_\Pi^2
{\cal J}} \label{op}
\end{equation}

\noindent By inserting this expression in the H-J-B equation
(\ref{H}), one obtains:
\begin{equation}
\partial_t{\cal J}+\mu_i S \partial_S{\cal J}+ {1 \over
2} \sigma_i^2 S^2 \partial_S^2 {\cal J} - {(\mu_i \partial_\Pi
{\cal J} +\sigma_i^2 S
\partial_{S\Pi} {\cal J})^2 \over 2 \sigma_i^2 \partial_\Pi^2
{\cal J}}=0 \label{PDE}
\end{equation}

\noindent The resolution of the above equation gives
$\phi(S,t)_{i+1}$ and the procedure can be restarted. We will now
apply this method up to order one. The form of (\ref{PDE}) and the
boundary conditions (\ref{bH}) suggest than $\cal  J$ takes the
following form:
\begin{eqnarray}
{\cal J}(t,S,\Pi)=  A(t,S)(\Pi+{\zeta \over 2})^2 -{\cal
G}^2-\zeta {\cal G}-{\zeta^2 \over 4}
\end{eqnarray}
\noindent with $\zeta$ such as $\partial_\zeta J=0$ and with the
boundary condition $A(T,S)=1$.

\noindent By inserting this expression in (\ref{PDE}) and
(\ref{op}), one then obtains the equations:

\begin{eqnarray}
\phi^*(u,t)=-S^{-1}( {\mu_i \over \sigma_i^2} + \partial_u
X)(\Pi+{\zeta \over 2})
\label{opt} \\
\partial_t X -(\mu_i+{1
\over 2}\sigma_i^2)  \partial_u X -{1 \over 2}\sigma_i^2(
\partial_u^2 X+(\partial_u X)^2)={\mu_i^2 \over \sigma_i^2}
\label{X}
\end{eqnarray}
\noindent with $u=ln(S)$ and $A(S,t)= e^{X(S,t)}$. The Lagrange
condition $\partial_\zeta J=0$ gives
\begin{equation}
\Pi + {\zeta \over 2}={{\cal G}-\Pi \over e^{X}-1} \label{lag}
\end{equation}

\noindent Let's define the parameter $\lambda \doteq {\mu_0^2
\over \sigma_0^2}$. In the zero-order approximation, the
volatility and return are constant and in this case, $A$ will be a
function of $t$ only given by
\begin{equation}
A(t,S)=e^{{\lambda}(t-T)}
\end{equation}
\noindent By inserting this expression in (\ref{opt}) and using
(\ref{lag}), one finds the optimal control (at zero-order):

\begin{equation}
\phi^*(S,t,\Pi)= {\lambda \over \mu_0}S^{-1}{ \Pi -{\cal G} \over
e^{{\lambda}(t-T)}-1}
\end{equation}
\noindent This expression allows to derive a stochastic equation
for the portfolio $\Pi$ subject to the condition $\Pi(0)=\Pi_0$:

\begin{equation}
{d\Pi \over (\Pi-{\cal G})}={1 \over (e^{\lambda(t-T)}-1)}(\lambda
dt + \sqrt{\lambda} dW)
\end{equation}

\noindent The mean of the portfolio $E[\Pi]$ is then given :
\begin{equation}
{d E[\Pi] \over dt}={\lambda}{ E[\Pi] -{\cal G} \over  e^{{
\lambda}(t-T)}-1}
\end{equation}
\noindent The solution is
\begin{equation}
E[\Pi]-{\cal G}=({\cal G}-\Pi_0){(1-e^{-{\lambda}(t-T)}) \over
(e^{\lambda T}-1)}
\end{equation}
\noindent One can then verify that the condition $E[\Pi_T]={\cal
G}$ is well satisfied. Finally, the optimal control $\phi^*_0$ at
order zero satisfies the following stochastic equation:

\begin{equation}
{d[\phi^*_0 S] \over \phi^*_0 S}= -\lambda t +{\sqrt{\lambda}
\over (e^{{\lambda}(t-T)}-1)} dW_t
\end{equation}
\noindent and the solution is
\begin{equation}
\phi^*_0= {\lambda \over \mu_0} S^{-1}{ ({\cal G}-\Pi_0) \over
(1-e^{-\lambda T})^{3 \over 2}} (1-e^{-\lambda (t-T)})^{1 \over 2}
e^{-\lambda t + {1 \over 2(e^{\lambda (t-T)}-1)}-{1 \over
2(e^{-\lambda T}-1)}+ \int_0^t {\sqrt{\lambda} \over (e^{\lambda
(t'-T)}-1)} dW_t'} \end{equation} \noindent As explained in the
first section, the hedging strategy depends explicity on the
history of the Brownian motion. To obtain a return and a
volatility that depend only on $S$ and $t$, we will take
$\bar{\phi_0}^*(S,t)=\phi^*(S,t,E[\Pi])$ as the demand of the
large trader:
\begin{equation}
\bar{\phi}_0^*(S,t)= {\lambda \over \mu_0} S^{-1} {({\cal G
}-\Pi_0)} { e^{-\lambda t} \over (1-e^{-\lambda T})}
\label{demand}\end{equation} We should note that the only models
considered by Merton are models of dynamical portfolio
optimization with no age effects \cite{merton}.

\noindent Let's derive now the one-order correction to this
hedging function. First, we give from (\ref{vola})-(\ref{reta})
${\mu^2 \over \sigma^2}$ up to the first order in $\epsilon$
(using (\ref{demand})):
\begin{eqnarray}
{\mu^2 \over \sigma^2}&=& {1 \over \sigma_0^2}( \mu_0^2 + 2
\epsilon (\partial_t \bar{\phi}^*_0 +{1 \over 2}\sigma_0^2S^2
\partial_S^2
\bar{\phi}^*_0 )) \\
&=&{1 \over \sigma_0^2}( \mu_0^2 + 2 \epsilon
\bar{\phi}^*_0(-\lambda+\sigma_0^2))
\end{eqnarray}
Then the equation (\ref{X}) at order one is
\begin{equation}
\partial_t X_1 -(\mu_0+{1
\over 2}\sigma_0^2)  \partial_u X_1 -{1 \over 2}\sigma_0^2
\partial_u^2 X_1={2 \bar{\phi}^*_0\over \sigma_0^2}
(-\lambda+\sigma_0^2)
\end{equation}

\noindent with boundary condition $X_1(0,u)=0$. The solution is
then given by
\begin{equation}
X_1(S,t)=2S^{-1}{({\cal G}-\Pi_0) \over (e^{\lambda T}-1)}{\lambda
\over \mu_0} { (\sigma_0^2-\lambda) \over (\mu_0-\lambda)
\sigma^2_0}(e^{\lambda (T-t)}-e^{\mu_0 (T-t)})
\end{equation}

\noindent and after a sightly long computation we find the
correction to $\phi_0^*$ which depends on $S$, $t$ and $\Pi$:

\begin{equation}
\phi^*(S,t,\Pi)={(\Pi-{\cal G}) S^{-1} \lambda \over \mu_0
(e^{\lambda(t-T)}-1)}(1+\epsilon \bar{\phi}^*_0
{(-\lambda+\sigma_0^2+\mu_0) \over \mu_0} -\epsilon {\mu_0 \over
\lambda} X_1-\epsilon {X_1 \over (1-e^{-\lambda(t-T)})})
\end{equation}

\noindent To obtain the next corrections for the implied
volatility and return, on should take $\phi^*(S,t,E[\Pi])$ as the
demand of the large trader. We have found that the large trader
strategy induces an implied volatility which is given at the
first-order by:

\begin{equation}
\sigma_{\mathrm implied}=\sigma_0(1-\epsilon \alpha e^{-\lambda t}
S^{-1}) +o(\epsilon^2) \label{im}
\end{equation}
with $\alpha$ a constant depending on the characteristic of the
portfolio ($\Pi_0$, $\cal G$, $T$, $\mu_0$, $\sigma_0$).


\section{Conclusion}
\noindent In this note, we have explained how to incorporate
easily  the effect of the hedging strategy in the market prices.
The derivative hedging gives a negative effect and the portfolio
optimization hedging
 a positive one. The main difficulty with
our optimization scheme was that the hedging function depends on
age effects, which was simply solved  by taking the mean over the
portfolio value for $S$ fixed. As explained the portfolio
optimization and the option pricing lead to relatively simple
non-linear partial differential equations (\ref{PDE})-(\ref{BS})
that can be numerically solved.

 \nonumsection{Acknowledgment} \noindent The author is
grateful to Stephane Marchand for careful reading and useful
discussion.

\renewcommand{\theequation}{\Alph{appendixc}.\arabic{equation}}
\appendix{: Modified Black-Scholes PDE}
\noindent The Black-Scholes analysis can be trivially modified  to
incorporate the hedging feedback effect. Let's $\Pi$ be the
portfolio and ${\cal C}(t,S_t)$ the value of the option at time
$t$:
\begin{equation}
\Pi=-{\cal C}(t,S_t)+\phi(t,S_t)S_t
\end{equation}
As a result of Ito's lemma, $d\Pi$ is  given by
\begin{eqnarray}
d\Pi&=&-d{\cal C}+\phi dS_t \\
&=&-(\partial_t {\cal C} + {1 \over 2} \sigma^2 S^2 \partial_S^2
{\cal C})dt+(\phi-\partial_S {\cal C})dS_t
\end{eqnarray}
A free-arbitrage condition gives $d\Pi= r(t) \Pi(t)$ with $r$ the
interest rate. The risk is therefore zero iff $\phi=\partial_S
{\cal C}$ which is the usual hedging position. The Black-Scholes
PDE (modulo adequate boundary conditions) is then
\begin{eqnarray}
&&\partial_t {\cal C} + {1 \over 2} { \sigma_0^2 \over (1-\epsilon
S\partial_S^2 {\cal C})^2} S^2 \partial_S^2 {\cal C}+r(-{\cal C}+S
\partial_S {\cal C})=0 \label{BS}
\end{eqnarray}
This non-linear PDE can be solved by expanding $\cal C$ (and so
$\phi$) as a formal sum in $\epsilon$. A numerical computation can
also be quite interesting and leads to the determination of an
implied volatility.

\end{document}